\documentclass[12pt]{article}
\usepackage[T1]{fontenc}
\usepackage[latin9]{inputenc}

\pdfoutput=1
\usepackage{graphicx}
\usepackage{graphics}
\usepackage{multirow}
\usepackage{underscore}

\title{Persistent storage of non-event data in the CMS databases}
\author{ M.De Gruttola$^{1,2,3}$, S.Di Guida$^{1}$, D.Futyan$^{4}$, F.Glege$^{2}$, \\
G.Govi$^{5}$,   V.Innocente$^{1}$, P.Paolucci$^{2}$, P.Picca$^{6,7}$, \\
A.Pierro$^{8}$, D.Schlatter$^{1}$,Z.Xie$^{9}$\\
\small $^1$CERN, Geneva, Switzerland,   \\       
\small$^2$INFN Sezione di Napoli, Naples, Italy, \\
\small$^3$Universit\`a  degli studi di Napoli ``Federico II'', Naples, Italy, \\
\small $^4$Imperial College, London, UK,\\
\small $^5$Department of Physics, Northeastern University, Boston, MA. 02115, USA,\\
\small $^6$Universit\`a di Milano - Bicocca, Dipartimento di Fisica G. Occhialini, Milan, Italy,\\
\small$^7$Centre de Calcul de l'IN2P3, CNRS, Lyon, France,\\
\small$^8$INFN Sezione di Bari, Bari, Italy, \\
\small$^9$Department of Physics, Princeton University, Princeton, NJ.08542, USA}
\begin{document}

\maketitle
\begin{center}
 E-mail: michele.de.gruttola@cern.ch\\
keywords: Software architectures, Computing 
\end{center}
\abstract{
In the CMS experiment, the non event data needed to set up the detector, or being produced by it, and needed to calibrate the physical responses of the detector itself are stored in ORACLE databases. 
The large amount of data to be stored, the number of clients involved and the performance requirements make the database system an essential service for the experiment to run. 
This note describes the CMS condition database architecture, the data-flow and PopCon, the tool built in order to populate the offline databases. 
Finally, the first experience obtained during the 2008 and 2009 cosmic data taking are presented.
}


\section{Introduction}
\label{sect:Introduction}

The large amount of data needed to set up the detector (tens of GBytes) and produced by it (few TBs per year) makes the database system a service which is essential for CMS data-taking operation.
During the construction phase, many databases, based on different technologies and software, were developed by the sub-projects in order to store all the detector and equipment data. 
In 2004, the CMS collaboration decided to start a common and central project, called the CMS Database Project, in order to converge towards an unique database technology and a set of central software tools supported by the experiment for all data taking.

The current layout of the database model and data-flow was developed after two workshops and one review and close interaction with the CMS subsystems.
 
The two most important requirements identified by CMS are:
\begin{itemize}
\item	CMS has to be able to operate without network connection between LHC Point 5 (P5) and the outside world (CERN network included). Therefore CMS must own an independent database structure based at P5.
\item	The offline condition data work-flow has to fit a multi-tier distributed structure as used for event data.
\end{itemize}
The Database Project group, with the help of IT and in collaboration with all the CMS sub-projects, designed a system based on 3 different databases, all based on the ORACLE technology (a commercial relational database system \cite{ORACLE}). 
\begin{itemize}
\item	 Online Master Database System ({\bf{OMDS}}) is the online master database located at P5 on the CMS online network. 
It stores the configuration of the detector and the non event data (condition data) produced by the sub-systems like slow control, electronics, data acquisition (DAQ) and trigger data. 
It is a purely relational database.
\item  Offline Reconstruction Condition database for ONline use ({\bf{ORCON}}), on the online network, stores all the offline condition data required online by the High Level Trigger (HLT) and offline for the event data reconstruction. 
It also contains conditions needed offline for data quality indication and for more detailed offline analysis. ORCON serves only as an intermediate storage of the latest offline condition data. The entire history of off line condition data is stored in ORCOFF.
The data contained in it are written using the POOL-ORA\cite{POOLORA} technology and are retrieved by the HLT programs as {\texttt{C++}} objects.  
\item  Offline Reconstruction Condition database for OFFline use ({\bf{ORCOFF}}) is the master offline database located at the Tier-0 site (CERN Meyrin) and it contains a copy of ORCON made through ORACLE streaming. ORCOFF
contains the entire history of all CMS condition data and serves prompt reconstruction as well as the condition deployment service to Tier-1/Tier-2 sides as input source.
Data contained in it are retrieved by the reconstruction algorithms as \texttt{C++} objects. 
\end{itemize}

\section{Non-Event Data Description}
\label{sect:CondDataDescription}

For each sub-detector, the non-event data to be stored in the CMS databases can be classified in different groups, according to their needs for meta-data (i.e., data to describe the data):
\begin{itemize}
\item{\bf Construction data} During the construction of the detector, data are gathered from both the production process and the produced items. 
Some of the construction data also belongs to the data types described below, and therefore were moved to the common data storage at the end of construction. 
The different CMS sub-detectors agreed to keep their construction data available for the lifetime of the experiment in order to be able to trace back production errors. 
The construction data and their storage will not be described in this document.
\item{\bf Equipment management data}  Detector items should be track-~ed in order to log their history of placements and repairs.
The classification of CERN as INB (Installation Nucleaire de Base \cite{INB}) requires, in addition, to keep a continuous trace of the location of irradiated items. 
Equipment management data contain, therefore, the location history of all items being installed at the experiment, in particular detector parts as well as off detector electronics.
Hence, the required meta-data must be time validity information. This data are stored in OMDS.
\item{\bf Configuration data} The data needed to bring the detector into any running mode are classified as configuration data. They comprise voltage settings of power supplies as well as programmable parameters for front-end electronics. 
Configuration data require a version and time validity information as meta-data. This data are stored in OMDS.
\item {\bf Condition data} The data describing the state of any detector subsystem are defined as condition data. These conditions are measured online and are stored in OMDS. They include data quality indicators such as bad channel lists and settings of the detectors needed offline (such as pedestals).  Condition data in OMDS are used in the online system for post mortem analysis of detector errors. Condition data needed for HLT and offline reconstruction are uploaded in ORCON, and must be described by a version and the time validity information corresponding to the set of data for which they are measured.  
\item {\bf Calibration data} The data describing the calibration and the alignment of the individual components of the different sub-dete-~ctors are labeled as calibration data.
These quantities (such as drift velocities, alignments constants, etc.) are evaluated by running dedicated algorithms offline. 
Since they are needed by HLT and for offline reconstruction, they appear only in the offline databases (ORCON and ORCOFF).
Calibrations must match the corresponding raw data coming from the collision events revealed by the detector.
Calibration data can be grouped by the version and the time range in which they are valid.
\end{itemize}

\section{The Database Architecture}
\label{sect:DBArchitecture}

Different data usage and access between online and offline determines the overall database architecture for the CMS experiment.
In the online network, data are mainly written into the database, so that the time for a database transaction to be committed is critical, while, in the offline network, data are mainly read from the databases.
Moreover, the online data are being written at random times, while the offline data must be synchronized with the events.
Since online data are used for error tracking, different data items must be accessible in order to be compared between each other; on the other hand, offline data must be grouped before they are read, so that they can be decoded according to predefined rules.

\begin{figure}[hbtp]
  \begin{center}
 \resizebox{1.0\textwidth}{!} {\includegraphics{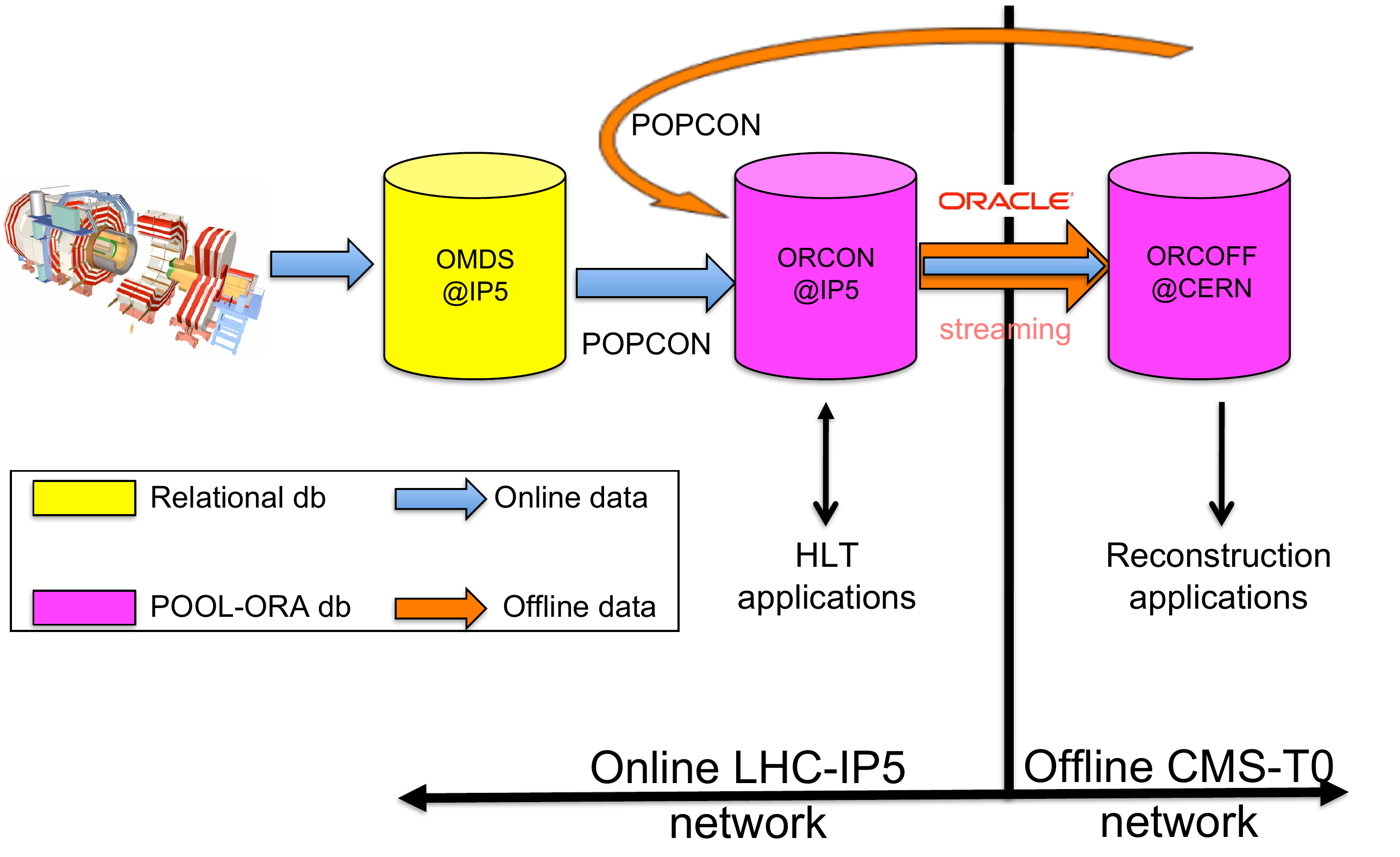}} \caption{Condition databases architecture.}
    \label{fig:CondDBArchitecture}
  \end{center}
\end{figure}

The general non-event data flow can be described as follows (see figure \ref{fig:CondDBArchitecture}): every subproject calculates and measures in advance all the parameters needed to setup its hardware devices, mainly related to the detector, DAQ and trigger. 
Hence, configuration data are prepared using the equipment management information, for both hardware and software. 
Different hardware setups can be stored at the same time in the configuration database, but only one will be loaded before the run starts.
During data taking, the detector produces many kind of conditions, to be stored in OMDS, from the slow control (PVSS\cite{PVSS}) and from other tools like DAQ (XDAQ), Run Control and data quality monitoring (DQM).
Part of OMDS data, needed by the HLT and offline reconstruction, will be transferred to ORCON.
A software application named PopCon (Populator of Condition Objects) operates the online to offline condition data transfer (the so-called O2O procedure), and encapsulates the data stored in relational databases as POOL-ORA objects. 
PopCon adds meta-data information to non-event data, so that they can be retrieved by both the HLT and the offline software. 

In addition to condition data transferred from OMDS to ORCON by the O2O procedure, calibration and alignment data determined offline are also written to ORCON, using again PopCon.
Finally, data are transferred to ORCOFF, which is the main condition database for the CMS Tier-0, using ORACLE streaming.
For massively parallel read-access, the ORCON and ORCOFF databases are interfaced with a 
cache system referred to as Frontier, which in case of ORCOFF is the mechanism used to distribute conditions data to the Tier-1 and Tier-2 centres outside CERN. 
Caching servers (squids)  are used to cache requested objects to avoid repeated access to the same data, significantly 
improving the performance and greatly reducing the load on the central database servers. Further details can be found in \cite{FRONTIER}.

As data taking proceeds, we can understand better and better how the detector works; therefore, this will require a new version of calibrations. 
When it will be available, it will be uploaded into ORCON using PopCon, and then streamed offline to ORCOFF.

\subsection{The Online Master Database}
\label{subsect:OnlineDB}

In the CMS experiment, the non event data needed to set up the detector, or being produced by the detector itself, is stored in OMDS.
The online database must allow for accessing individual, ideally self explaining data items: hence a pure ORACLE access and manipulation structure has been chosen for OMDS.

The data size is expected to become very large (several TBs), and, since condition data will constantly flow into the database, the time needed to store these data in OMDS is a critical issue. 
To fulfill these requirements, each sub-detector has designed its own database schema, reflecting as far as possible the detector structure.

The total amount of data stored in OMDS is about 1.5 TB in 100 days of data taking. 
This rate was extrapolated from the 2008 and 2009 cosmic runs: in particular, in the account on OMDS for non-event data coming from the electromagnetic calorimeter (ECAL), about 5 GB of data per day are stored.

\subsection{Offline database}
\label{subsect:OfflineDB}

As shown in figure \ref{fig:CondDBArchitecture}, the CMS database infrastructure envisages two offline databases intended for condition data:
\begin{itemize} 
\item {\bf ORCON} is a part of the database infrastructure at P5. 
It contains CMS condition and calibration data and serves as an input source to the HLT. It is also an intermediate storage element of the latest condition objects.      
\item {\bf ORCOFF} is the offline master database for condition data. 
It is located at the Tier-0 (CERN Meyrin). It contains a copy of the data in ORCON, and the entire history of all CMS non-event data. It serves as an input source for all offline reconstruction. The data in ORCOFF are accessed from the Tier-0, and from all Tier-1 and Tier-2 centres via the Frontier caching mechanism.
\end{itemize}

ORCON possess identical ``schemas'' as ORCOFF, optimized for the offline usage. 

Together with the production databases, CMS users can also use a ``development'' and an ``integration'' database, intended for tests, and accessible from the offline network: 
\begin{itemize}
\item{\bf ORCOFF\_PREP}, offline database for preparation and development purposes. 
\item{\bf ORCOFF\_INT}, offline database for integration. It should be used if all the tests on ORCOFF_PREP have been successful. 
\end{itemize}

Since the 2009 cosmic data taking (namely, CRAFT2009), CMS deployed a ``development'' and an ``integration'' database also in the online network:
\begin{itemize}
\item{\bf cmsdevr}, offline database in the online network for preparation and development purposes. 
\item{\bf cmsintr}, offline database in the online network for integration. 
It should be used if all the tests on cmsdevr have been successful. 
\end{itemize}

The data access (both insertion and retrieval) is controlled only by the {\texttt{C++}} based POOL-ORA API (see \ref{subsect:POOL-ORA}). 
In the offline databases, only a subset of configuration data and condition data, as well as all calibration data, must be stored.
All these data need a tag, labeling their version, and an interval of validity for describing their time information, as meta-data. 
The interval of validity (IOV \cite{CondCHEP09}) is the contiguous (in time) set of events for which non-event data are to be used in reconstruction. 
According to the use-case, the IOV will be defined in terms either of GPS-time (mainly for condition data) or ``run-number'' range (usually for calibrations). 
Whilst the IOV for some electronic related conditions (e.g. pedestals and noises) is identical to the time interval in which these data were used in the online operations, some calibration data may possess an IOV different from the time range in which they were defined.
For this reason, the IOV assignment for a given set of condition data is carried out at the offline level.
Each payload object, i.e. each data stored as a POOL-ORA object in ORCOFF, is indexed by its IOV and a tag, a label describing the calibration version, while the data themselves do not contain any time validity information.     

The matching with the raw data from the collision events is indeed possible via these meta-data: the reconstruction algorithms for the analysis of a given run query the offline condition data corresponding to the same run grouped through a set of tags, called \emph{global tag \cite{CondCHEP09}}.  

The policy established by the Database Project for the CMS community is to write any condition/calibration data in ORCON; the data are then copied to ORCOFF using the ORACLE streaming tool.

The size of condition data stored in ORCON and ORCOFF, where only a subset of condition data will be uploaded, is decreased by a factor of 20 with respect to OMDS. This is a great success of the entire architecture.

In section \ref{subsect:POPCON} the PopCon framework is described, while in section \ref{susect:CentralPopulation} more information about the online-to-offline (O2O) transfer operated by PopCon is given.

\subsection{POOL Object Relational Database Access: POOL-ORA}
\label{subsect:POOL-ORA}

POOL\cite{POOL}, the persistency framework for object storage for the LHC experiments, was successfully used by ATLAS, CMS and LHCb to handle data during data challenges. 
The relational back-end of POOL, namely POOL-ORA, is chosen by the CMS experiment for condition data handling. 
The implementation sits on top of a generic relational database layer. 
The POOL-ORA interface used for handling non-event data is identical to that of POOL-ROOT used  for handling event data. 
The main feature is that the transient object model drives the database model: the designers of the offline data model do not need to know the tabular representation of the data in the database. 
The offline database schema is automatically created from the definitions of the persistent-capable objects, by following the Object Relational Mapping (ORM) rules. 
The data are retrieved from the underlying relational database, then materialized as \texttt{C++} objects in memory by following the dictionary information, hence finding the corresponding entries in the ORM files.

\begin{figure}[hbtp]
  \begin{center}
    \resizebox{0.7\textwidth}{!}{\includegraphics{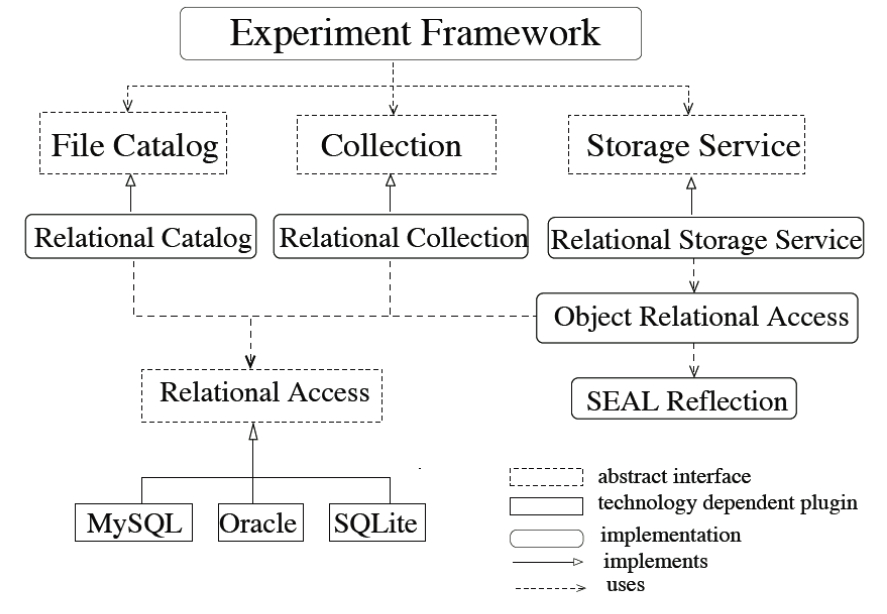}} \caption{POOL-ORA in the overall architecture.}
    \label{fig:POOLORA}
  \end{center}
\end{figure}

As shown in figure \ref{fig:POOLORA}, PoolORA consists of three domains:
\begin{itemize}
\item {\bf COmmon Relational Access Layer (CORAL)} defines a vendor independent API for relational database access, data and schema manipulation. 
Three technology plugins, for ORACLE, MySQL and SQLite technologies, are released together with the API. 
\item {\bf Object Relational Access (ORA)} implements the object relational mapping mechanism, and is responsible for transforming \texttt{C++} object definitions into relational structures and vice-versa.
\item {\bf Relational Storage Service} implements the POOL Storage Service using the relational Access and the Object Relational Access components. 
\end{itemize} 

CMS contributes to the development of the POOL-ORA infrastructure in order to ensure that it satisfies the requirements from the CMS community.

\subsection{PopCon}
\label{subsect:POPCON}

PopCon \cite{PopCon} is a mini-framework within the CMS software CMSSW\cite{CMSSW} that transfers the condition objects from a user-defined data source to the offline databases.

Popcon is based on the \texttt{cmsRun} infrastructure of the CMS software framework\cite{EDM}. It is possible to use different data sources such as databases, ROOT files, ASCII files, and so on. 
A {\texttt C++} object type (built in type, structure, class, template class) which contains the non event data, must be defined into the CMS software framework. For each condition object (payload) class a PopCon application is created.

\begin{figure}[hbtp]
  \begin{center}
    \resizebox{1.0\textwidth}{!}{\includegraphics{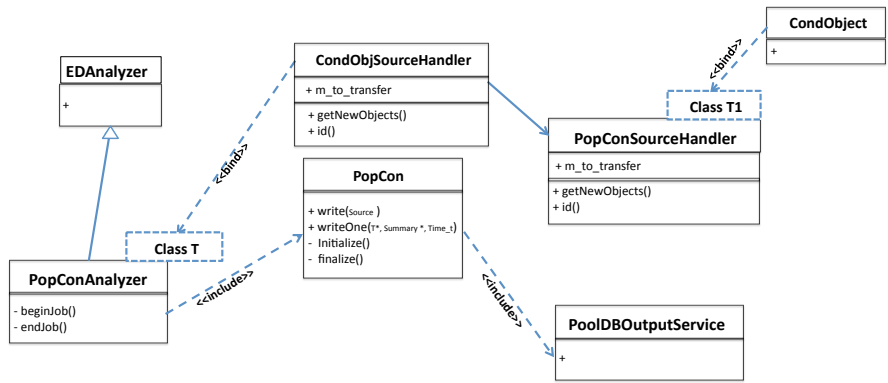}}\caption{Schema of the classes for the PopCon package.}
    \label{fig:PopConSchema}
  \end{center}
\end{figure}

The core framework consists of three classes (two of them are {\texttt C++} templates), as can be seen in figure \ref{fig:PopConSchema}:

\begin{itemize}
\item PopCon
\item PopConSourceHandler
\item PopConAnalyzer 
\end{itemize}

Once the {\texttt C++} condition object is embedded into CMSSW, the ``detector user'' provides the code which handles the data source and specifies the destination for the data, writing a derived class of PopConSourceHandler, where all the online (source handling) code goes. 
The user instantiates the objects, provides the IOV information for such objects and configures the database output module. 
The PopCon configuration file associates the tag name defined according to some specific rules, to the condition object. 
Once the object is built, the PopCon application writes the data to the specified account in the offline database. 
Sub-detector code does not access the target output database: it only passes the objects to the output module.

The analyzer object holds the source handling object. It also serves to implement  some additional functionality such as:
\begin{itemize}
 \item Locking mechanism.
 \item Transfer logging.
 \item Payload verification (IOV sequence).
 \item Application state management.
 \item Database output service.
\end{itemize}
The writer in PopCon iterates over the container of user objects and stores it in the user-configured data destination. 

Any transaction towards ORCON is logged by PopCon, and the process information is sent to a database account as well. A monitoring tool for this information was developed, in order to check the correctness of the various transactions, and to keep trace of every upload for condition data, see section \ref{subsect:PopConMonitoring}.

\section{First Experience in Operating the Population on the Condition Databases in 2008 and 2009}
\label{sect:DBPopulation}

In the 2008 and 2009 global runs (with and without the magnetic field), the great majority of the condition data was transferred offline using a PopCon application. 
Great effort was devoted by the CMS database project team to the integration of all the software and to the infrastructural chain used to upload the calibration constants into the CMS condition databases. 
Many tools were provided to help the sub-detector responsible people to populate their database accounts. 
A central procedure, based on an automatic up-loader into ORCON on a dedicated machine in the online network, was successfully deployed during 2008, and became the recommended way to populate ORCON during 2009 data taking.

\subsection{Condition Objects Written Using PopCon in 2008}
\label{subsect:2008CondObjects}

As stated before, each piece of condition data (pedestals, Lorentz angles, drift time, etc.) corresponds to a {\texttt C++} object (``CondObjects'') in the CMS software. 
Each object is associated with a PopCon application which writes the payload into ORCON. 
Table \ref{tab:pageLayout} lists all the CondObjects used in 2008, grouped according to the subsystem they belong to. 
For each object the type, the approximate data size in ORCON and the upload frequency are also reported.  
   
 \begin{table}[htbp]
    \caption{2008 CMS condition objects list}
    \label{tab:pageLayout}
    \begin{center}
      \begin{tabular}{||c|c|c|c|c||} \hline 
            Subsystem   & Name & Type & Data size & Frequency\\  \hline
  \multirow{3}{*} {Pixel} & FedCablingMap & online configuration& 1K& once (before the run )\\        
 & LorentzAngle & offline calibration & 1MB & each run (if different) \\ 
& CalibConfiguration  & online calibrations & 5KB & each calibration run \\ \hline
  \multirow{5}{*} {Tracker} & FedCabling & online configuration& 1K&once \\        
& BadStrip & online condition & 1MB&  each run (if different)\\ 
& Threshold & offline calibration & 1MB&  each run (if different)\\        & Pedestals & offline calibration & 1MB&  each run (if different)\\  
& Noise & offline calibration & 1MB&  each run (if different)            
\\ \hline

  \multirow{2}{*} {Ecal} & Pedestals & online calibration& 2MB& daily \\     & LaserAPDPNRatios & online calibration & 2MB& hourly        
\\ \hline

 \multirow{5}{*} {Hcal} & ElectronicsMap & online configurations & 1MB& once (before the run)  \\  
& Gains & offline calibrations & 1MB& each run \\  
& Pedestals & offline calibrations & 1MB& each run \\  
& PedestalsWidths & offline calibrations & 1MB& each run \\  
& QIEData & online calibrations & 1MB& each run 

\\ \hline

 \multirow{7}{*} {CSC} & ChamberMap & online configuration & 10KB & monthly  \\  
& CrateMap & online configuration & 10KB & monthly  \\  
& DDUMap & online configuration & 10KB & monthly  \\  
& ChamberIndex & online configuration & 10KB & monthly  \\  
& Gains & offline calibrations & 2MB& each run  \\  
& NoiseMatrix & offline calibrations & 2MB& each run  \\  
& Pedestals & offline calibrations & 2MB& each run  \\  
\hline

\multirow{5}{*} {DT} & ReadOut & online configuration & 10MB & once  \\  
 & CCBConfig &  online configuration  & 100KB & once (before the run)   \\
 & T0 & offline calibration & 10MB & rare  \\
 & TTrig & offline calibration & 1MB & at trigger change   \\
 & MTime & offline calibration & 1MB & daily   \\
\hline

\multirow{3}{*} {RPC} & EMap & online configuration & 10MB & once  \\
& L1Config & online configuration & 10MB & once  \\
& Cond & online conditions & 10MB & daily  \\
\hline

\multirow{1}{*} {DAQ} & RunSummary & run conditions & 10KB & run start/end  \\
\hline
      \end{tabular}
    \end{center}
  \end{table}

\subsection{Central Population of the Condition Data-~bases}
\label{susect:CentralPopulation}

A central procedure was set up in 2008, and used since then, for populating the CMS condition databases: it exploits a central account, explicitly devoted to the deployment of tools and services for the condition databases, in the CMS online network. 
On that account, a set of automatic jobs was centrally set up for any single sub-detector user, in order to both populate ORCON and monitor any transactions to it. 

Two possibilities are given to users:
\begin{enumerate}
\item to run automatically the application that reads from any online source, assigns tag and interval of validity, and uploads the constants into ORCON (mainly for condition data). 
The time interval of the automatic jobs is negotiated with the users;  
\item to use the account as a drop-box: users copy the calibrations in the SQLite\cite{SQLite} format into a dedicated folder for each sub-detector, and then these data are automatically exported in ORCON (mainly for offline calibration data).  
\end{enumerate}

\begin{figure}[hbtp]
  \begin{center}
    \resizebox{1.0\textwidth}{!}{\includegraphics{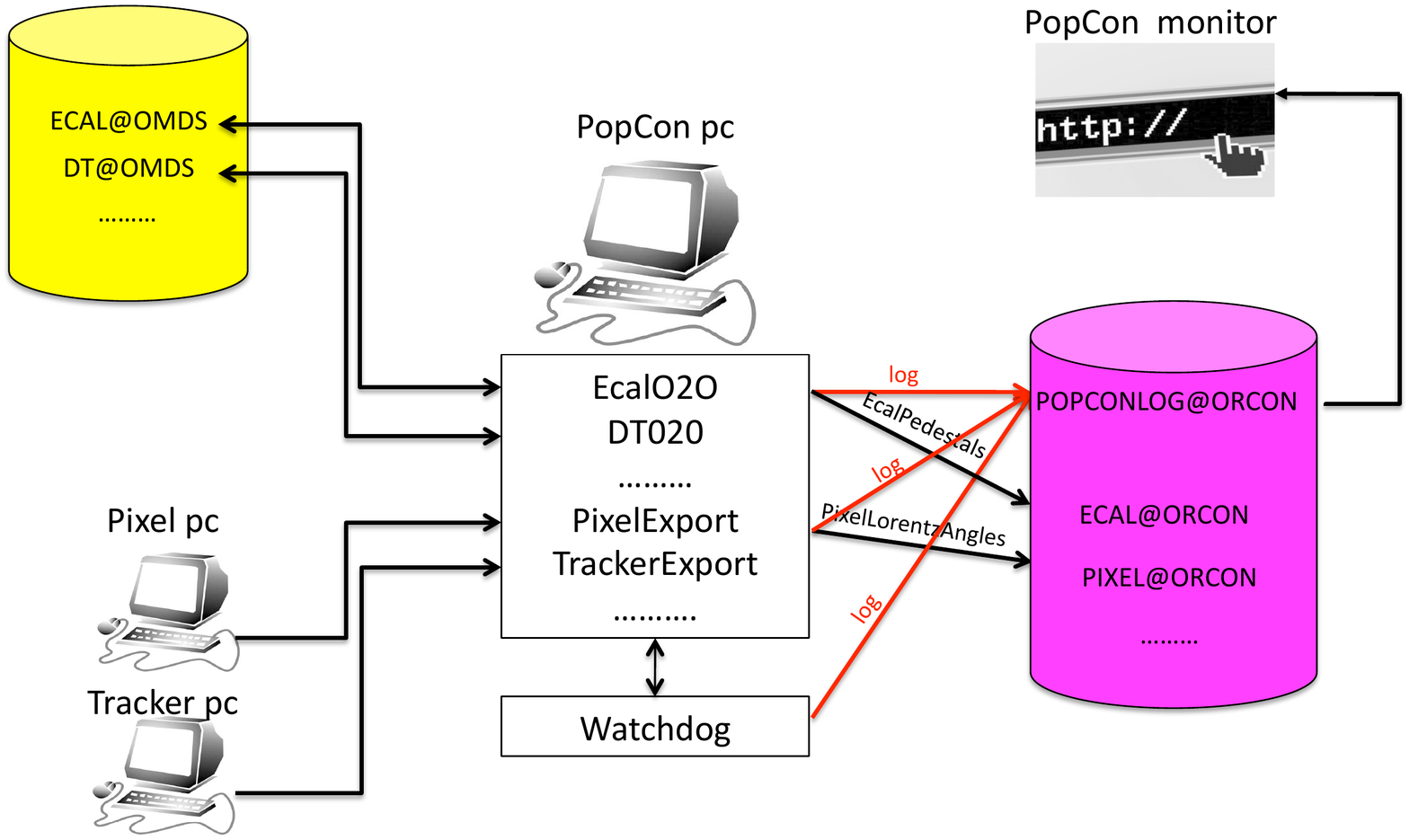}} \caption{Schematic illustration of the central system used to populate ORCON, and of the web monitoring system.}
    \label{fig:DropBox}
  \end{center}
\end{figure}

Figure \ref{fig:DropBox} shows a sketch of the central system used to populate the condition database. 
Each sub-detector responsible person may transfer the payload onto the central PopCon PC, that then automatically manages the exportation into the ORCON database (using a specific set of Subdetector Export scripts). 
Other automatic scripts (e.g. ECALO2O, DTO2O...) check to see if new conditions have appeared in the online table, and, if so, perform the data transfer from OMDS to ORCON.   
The PopCon applications transfer each payload into the corresponding account, and create some log information which are subsequently stored in the PopConLog account on ORCON itself. 

Each automatic job is associated with a ``watchdog'' tool that monitors its status.
The job monitoring information are also logged into the PopConLog account on ORCON.  

Since 2009 data taking a drop-box in the offline network has recently also been deployed. This infrastructure, using web applications inside Virtual Machine technology, and the Python programming language performs an automatic exportation of the source data to ORCON.  This is important in order to allow automation of offline calibration and alignment procedures, by eliminating the need for those procedures to connect to the online network.

\subsection{PopCon Web Based Monitoring}
\label{subsect:PopConMonitoring}

A dedicated web based application, {\it PopCon monitoring} \cite{PopConMonitoring}  was set up on a CMS web server in order to look at all the logged information, hence monitoring the activity on the condition databases. 

\begin{figure}[htbp]
\begin{center}
\resizebox{0.8\textwidth}{!}{\includegraphics{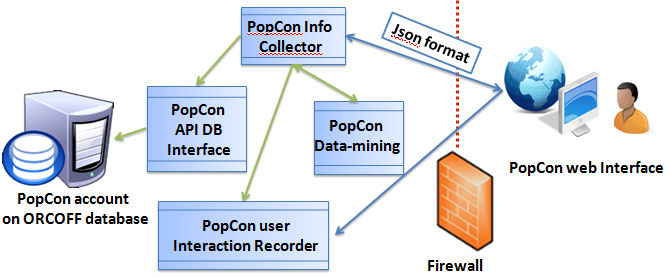}}
\end{center}
\caption{PopCon monitoring Architecture} \label{fig:PopConGuiArchitecture}
\end{figure}

\begin{figure}[htbp]
\resizebox{1.0\textwidth}{!}{\includegraphics{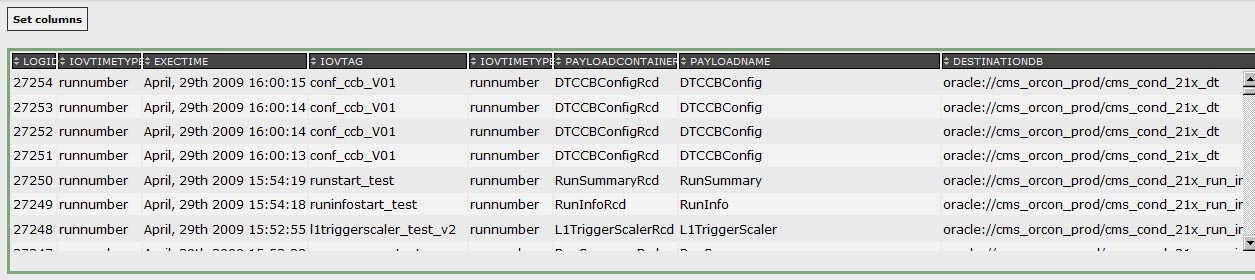}}
\caption{\label{fig:popConTable} 
The PopCon web interface represents information about database transactions in 
different types: both charts and tables.
A user can easily add or remove columns by clicking the checkbox
and columns can also be sorted. Information could be grouped according to different filters.
}
\end{figure}

\begin{figure}[hbtp]
  \begin{center}
    \resizebox{1.0\textwidth}{!}{\includegraphics{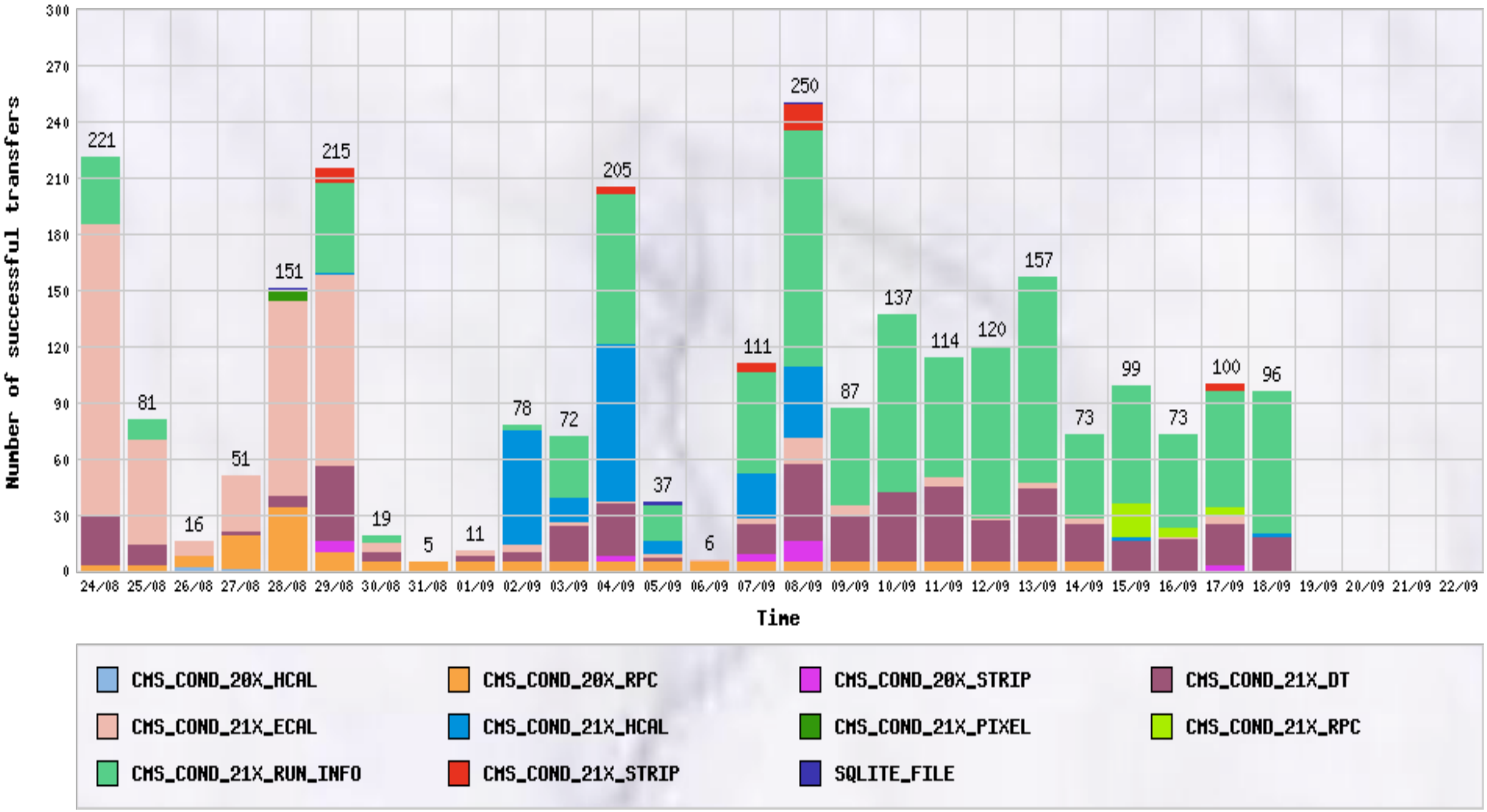}} \caption{PopCon activity between end September-beginning of October 2008.}
    \label{fig:PopConAct}
  \end{center}
\end{figure}

\emph{PopCon monitoring} comprises five main layers (see figure \ref{fig:PopConGuiArchitecture}):

\begin{itemize}
\item the \textbf{PopCon API DB Interface} retrieves the
entities monitored by PopCon;
\item the \textbf{PopCon user Interaction Recorder} is a collection 
that retains a history of interactions with each user.
\item the \textbf{PopCon data-mining} extracts patterns from data, 
entities monitored by PopCon. 
and the history of recorded user interactions, 
hence transforming them into information such as warnings, errors or alarms according to use case models.
\item  the \textbf{PopCon info collector} aggregates the
information produced by the different database transactions 
and the history of recorded user interactions, and encodes them in JSON\footnote{JSON (JavaScript Object Notation \cite{JSON}) is a lightweight data-interchange format.} format.
\item the \textbf{PopCon Web Interface} displays the information about the database transactions 
from the different user perspectives,
organizing data in tables (see figure \ref{fig:popConTable}) and/or charts (see figure \ref{fig:PopConAct}).
\end{itemize}

\begin{figure}[hbtp]
\resizebox{1.0\textwidth}{!}{\includegraphics{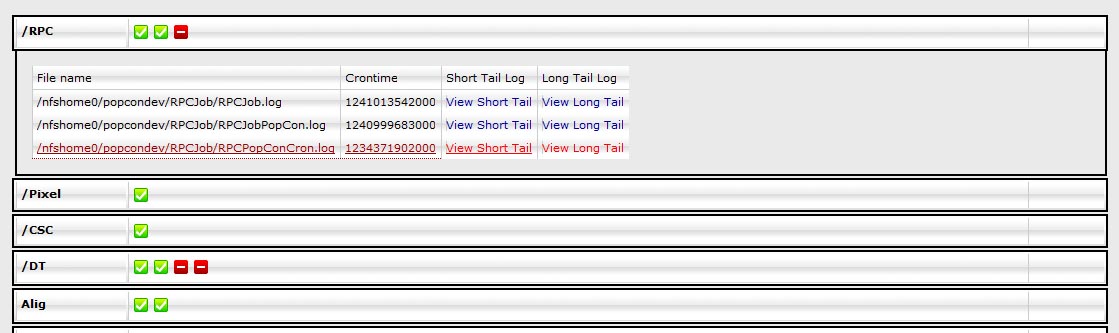}} \caption{Screenshot of the web page produced by the monitoring system that checks the watchdog tools for the automatic population of ORCON. Different colours helps to identify, quickly, the seriousness of the problem.}
  \label{fig:WatchDog}
\end{figure}

In addition, two other web pages, very useful for database transaction monitoring, are produced:
\begin{enumerate}
\item an activity summary, in which the number of ORCON transactions, the subsystem involved, the IOV and tag can be displayed, according to the users' requests. An example is shown in figure \ref{fig:PopConAct}. 
\item the logs of all the central scripts, as produced by the watchdog tools. 
Looking at those logs, the correct behaviour of the central uploader machine is controlled, so that an alarm system, based on that information, can recognize if some exports were not successful and, eventually, inform the end-user of the error occurred. 
A screenshot of the page is shown in figure~\ref{fig:WatchDog}.    
\end{enumerate}

Figure \ref{fig:PopConAct} reports all the transactions towards the condition database accounts occurring in a month of cosmic data taking in 2008. 
As the summary plot points out, almost all sub-detectors used PopCon to upload calibration constants to the condition databases.
An average of one hundred PopCon applications per day were run during the test runs in Summer/Fall 2008, hence one hundred connections per day to the condition databases took place.

During the whole year 2008 commissioning exercises, the total amount of condition data written in the production database was approximatively 1 TB. 
The same rates and volumes are expected for the future LHC collision data.
Moreover, no network problems, neither for the online-offline streaming, nor for Frontier were detected. 
All the conditions and calibrations were properly evaluated for CRUZET and CRAFT data taken in 2008, and are being evaluated in current 2009 CRAFT runs, leading to several global tags used for the reconstruction and the analysis of the cosmic ray data by the whole CMS community.

\section{Conclusions}
\label{sect:Conclusions}

A database system has been set up in order to upload, store and retrieve all non-event data for the CMS experiment. 
The system relies on ORACLE database for data storage, and on the POOL-ORA technology for the HLT and the offline algorithms. 
An application called PopCon, fully integrated into the CMS software framework CMSSW, operates the population of the offline databases, and adds some additional services such as transaction logging, hence allowing monitoring of all transactions against the account dedicated to each subdetector in the offline databases.  
The whole chain was deployed and tested succesfully during 2008 challenges with cosmic rays, and was further improved and upgraded for 2009 cosmic ray data taking: these tests demonstrate that the system we have just described is stable and robust enough for the 2009-2010 collision data taking.

\end{document}